\documentclass[aps,epsf,preprint]{revtex4}
\usepackage{amssymb}

\def\1{{\bf 1}}
\def\[{\left[}
\def\]{\right]}
\def\be{\begin{eqnarray}}
\def\ee{\end{eqnarray}}
\def\bm{\begin{pmatrix}}
\def\em{\end{pmatrix}}
\def\nn{\nonumber}
\def\({\left(}
\def\){\right)}
\def\bk#1{\langle#1\rangle}
\def\eq#1{(\ref{#1})}

\def\a{\alpha}

\def\s{\sigma}

\def\f{\phi}

\def\G{{\cal G}}

\def\x{\times}

\def\d{\delta}

\def\labels#1{\label{#1}}
\def\edc{\end{document}}

\def\bn{\begin{enumerate}}
\def\i{\item}
\def\en{\end{enumerate}}

\def\rd{\sqrt{2}}
\def\rt{\sqrt{3}}

\def\rs{\sqrt{6}}
\def\diag{{\rm diag}}

\def\ba{\begin{array}}
\def\ea{\end{array}}
\def\bc{\begin{center}}
\def\ec{\end{center}}

\def\edoc{\end{document}}

\def\^{$\wedge$}

\begin{document}

\title{A Built-in  Horizontal Symmetry of $SO(10)$}
\author{C.S. Lam}
\address{Department of Physics, McGill University\\
 Montreal, Q.C., Canada H3A 2T8\\
and\\
Department of Physics and Astronomy, University of British Columbia,  Vancouver, BC, Canada V6T 1Z1 \\
Email: Lam@physics.mcgill.ca}

\begin{abstract}
In a renormalizable $SO(10)$ theory, all fermion mass matrices are linear combinations of three fundamental types,  $M^{10}, M^{\overline{126}}$, and $M^{120}$, whose superscripts indicate their $SO(10)$ transformation properties. We point out
that each of these fundamental mass matrices possesses a natural symmetry that can be used to generate
an unbroken horizontal symmetry  $\G$, if the natural symmetry is taken to be the residual symmetry.
This built-in symmetry is a Coxeter group.
 If  it  is finite, it must be 
one of  five groups,
 $S_4,\  Z_2\x S_4$,\  $Z_2\x A_5$, plus two  `rank-4' groups. These symmetries place constraints on the fundamental
mass matrices and 
reduce the number of parameters 
in an SO(10) fit.  Since they are built-in and can be derived theoretically, it is 
 hoped that they impose better constraints than those without a theoretical basis, but that is to be confirmed because
there is  no attempt to fit the experimental data in this article, except to count the number of free parameters.
To illustrate the similarities and differences of various kinds of  constraints,
a comparison is made with an existing $S_4$ model, and with models possessing the Fritzsch 
texture. 
\end{abstract}
\narrowtext
\maketitle
\section{Introduction}
The 12 fundamental fermions in nature are divided into three generations. Those in the same  differ from one another by their Standard-Model quantum numbers, but there are no quantum numbers  to tell the generations apart. This asymmetry,
and the difficulty in identifying a horizontal symmetry, is partially due to its breaking, needed to 
generate mixing and to keep the masses of different generations different. Under such circumstances, not only the horizontal symmetry group has to be identified, it is also
necessary to know how to break it. There are
many strategies used 
to deal with such a task \cite{hs},  one of them is to identify the natural symmetry found in the leptonic mass matrices
with the residual horizontal symmetry left over after  breaking. With that assumption, the unbroken
horizontal symmetry  can be generated from the natural symmetry \cite{lam}. In the neutrino sector, this natural
symmetry is $Z_2$, or $Z_2\x Z_2$. In the charged-lepton section, it is $Z_k$ for an arbitrary $k>2$.

Unfortunately, the horizontal symmetry for leptons \cite{lepton} so obtained is very different from the symmetry for quarks \cite{quark}
obtained in a similar manner. It is hard to reconcile the small mixing angles of quarks with the generally large mixing angles of neutrinos.

To ensure a common origin of symmetry for both leptons and quarks, 
 a Grand Unified Theory (GUT) is called for. In what follows we shall take that to be
$SO(10)$ \cite{so10, so10hs, moh,ft}, whose irreducible representation ${\bf 16}$ accommodates all left-handed single-colored
fermions in one generation, including the heavy
Majorana neutrino implicated in type-I seesaw and leptogenesis. Since ${\bf 16\x 16=10+\overline{126}+120}$,
every fermion mass matrix in a renormalizable theory is a linear combination of  three types of 
fundamental mass matrices, $M^{10}, M^{\overline{126}}$,
and $M^{120}$, whose superscripts indicate their $SO(10)$ transformation property.
It turns out that $M^{10}$ and $M^{\overline{126}}$ are symmetric matrices 
and $M^{120}$ is antisymmetric. If $a,b=1,2,3$ are the generation indices, then $M_{ab}=\pm M_{ba}$ is a relation between
generation $a$ and generation $b$, thus akin to a horizontal symmetry. 
We shall show in the next section that indeed every fundamental mass matrix has a {\it natural symmetry} $(Z_2)^n$, the direct
product of $n$ $Z_2$'s, with some $n$ between 1 and 7. if we identify them as {\it residual symmetries}, then
they can be used to generate an unbroken horizontal group $\G$. 

It will be shown in the next section that this built-in horizontal symmetry   $\G$ is a Coxeter group. Moreover, if it is
 finite, then it must be one of five groups. The origin of this strong result is
our insistence that the residual symmetry left over after breaking is the natural symmetry $(Z_2)^n$, a requirement that is not
always obeyed in existing models \cite{so10hs, moh}.

In the usual approach of a renormalizable $SO(10)$ theory, there are three Higgs fields $\f^a\ (a=10, \overline{126}, 120)$ in the Yukawa terms.  Vacuum expectations $\bk{\f^a}$ are assigned from which the
fundamental mass matrices $M^a$ are computed. If a horizontal  symmetry $\G$ is present, 
then the residual symmetry ${\cal R}^a$ of $M^a$ is
generated by elements $g\in\G$ such that $g^T\bk{\phi^a}g=\bk{\f^a}$. In the present bottom-up approach, ${\cal R}^a=(Z_2)^{n_a}$
is given by the natural horizontal symmetries of $SO(10)$,  $\G$ is generated by these ${\cal R}^a$'s, thus the equivalent
vacuum alignments are invariant eigenvectors of some {\it order-2} elements of $\G$.

This natural symmetry in $SO(10)$ is reminiscent of the 
natural symmetry $Z_2$, or $Z_2\x Z_2$,
of the neutrino mass matrix \cite{lam}.  These two cases have indeed the same origin, arising from the symmetric nature
of the neutrino mass matrix on the one hand, and the symmetric or antisymmetric nature of the 
$SO(10)$ fundamental mass matrices on the other. 
 In the leptonic case, the horizontal group is generated by the $(Z_2)^n$ residual symmetry in the 
neutrino sector, with $n=1$ or 2, and a $Z_k$ residual symmetry in the charged-lepton sector, with $k>2$ quite arbitrary. Together they can generate
an infinite number of fairly complicated finite groups 
that have three-dimensional irreducible representations (3dIR) appropriate to the
three generations. One must comb through all of them \cite{hs, lam, lepton, quark}  to fish out those whose neutrino-mixing predictions
agree with data. In the $SO(10)$
case, the residual symmetry of every fundamental mass matrix is of the type $(Z_2)^n$, without any $Z_k$ for $k>2$.
As a result, all
 the finite horizontal symmetry groups $\G$
they can generate  are known, and among them only five possess 3dIR. This makes the search of a finite horizontal symmetry 
for quarks and leptons together in $SO(10)$ much simpler than for leptons alone, though this simplicity is marred by
the complexity of having to verify  the validity of the vertical symmetry $SO(10)$ at the same time.

Note that this derivation of a built-in horizontal symmetry for $SO(10)$ relies on the symmetric or antisymmetric nature
of the fundamental mass matrices, which comes about  partly because all the fermions are contained in a {\it single}
 representation {\bf 16}, so
it would work in so straight forward a manner if the GUT was $SU(5)$.

The natural symmetry of $SO(10)$ together
with the property of Coxeter groups   will be discussed in Sec.~2.
Their 3dIR will be presented in Sec.~3. The constraints a
symmetry puts on the fundamental mass matrices $M$ will be given in Sec~4. Its application to
$SO(10)$  to determine the fermion masses and mixings will be discussed in Sec.~5, including
horizontal symmetry constraints on the fermion mass matrices and  resulting number of real parameters.
 In Sec.~6, a comparison with an existing $S_4$ model \cite{moh} is made, and also 
a comparison with models \cite{ft} possessing the Fritzsch texture \cite{fr}, to illustrate
the similarities and differences of various constraints.  A summary is presented in Sec.~7
to conclude the article.

\section{Natural Symmetry and Coxeter Groups}

Suppose $M$ is a symmetric matrix, with non-degenerate 
eigenvalues $m_i$, and normalized eigenvectors $u_i$. By
studying the matrix element $u_j^TMu_i=m_ju_j^Tu_i=m_iu_j^Tu_i$, 
it is easy to see that $u_j^Tu_i=0$ if $i \not=j$, hence $u_j^Tu_i=\d_{ij}$, and
$M$ can be written in the dyadic form $M=\sum_i m_iu_iu_i^T$. Define $s=\sum_i \s_iu_iu_i^T$ with some unknown
$\s_i$. Then
$s=s^T$, and $s^TMs=\sum\s_i^2m_iu_iu_i^T=M$ if and only if $\s_i^2=1$ for all $i$. Such an $s$ obeys $s^2=1$,
and is a symmetry of $M$. Since each of the three $\s_i$'s can be either $+1$ or $-1$, there are 8 possibilities,
with one  being the identity matrix. These $s$'s thus generate a residual symmetry group $(Z_2)^n$, with 
$n$ between 1 and 7.

If $M$ is antisymmetric, then $u_j^TMu_i=m_iu_j^Tu_i=-m_ju_j^Tu_i$ tells us that
 the non-zero eigenvalues come in opposite pairs,
$(m_i,-m_i)$. It is therefore convenient to divide the index $i$ into two groups, with  $-a$ and $a$ 
labeling  the non-zero eigenvalues, so that $-m_a=m_{-a}$,
and   $A$ labeling the zero eigenvalues. 
In that case the orthonormal relations of the eigenvectors become $u_a^Tu_b=\d_{a,-b},\  u_A^Tu_B=\d_{AB}$,
and $u^T_{\pm a}u_A=0=u^T_Au_{\pm a}$. The dyadic form of $M$ is then
$M=\sum_{i=\pm a} m_iu_iu_{-i}^T$. Let $s=\sum_{i=\pm a} \s_iu_iu_{-i}^T+\sum_{A}\s_Au_Au^T_A$, with $\s_a=+\s_{-a}$. Then $s=s^T$, and $s^TMs=
\sum_{i=\pm a} \s_i^2m_iu_iu_{-i}^T=M$ if and only if $\s_{\pm a}^2=1$. There is no restriction on $\s_A$ but we will 
choose them
to be either $+1$ or $-1$, so that once again $s^2=1$.
For $3\x 3$ matrices, there is only one $a$ and one $A$, with $\s_a=\s_{-a}=+1$ or $-1$, and $\s_A=+1$ or $-1$. Hence
the residual symmetry group of antisymmetric matrices is $(Z_2)^n$, with $n=1, 2$. 

Thus for each fundamental mass matrix $M$ which is either symmetric or antisymmetric, 
one or more operators $s=s^T$ with $s^2=1$ can be found so
that $s^TMs=M$. If we identify the natural symmetry with the residual symmetry after breaking,
then the minimal  unbroken horizontal symmetry group 
is the group generated by all these distinct $s$'s. Let us use a subscript to distinguish these generators,
and proceed to find the structure of the group.
Suppose $s_bs_c$ has an order $o_{bc}$, so that $(s_bs_c)^{o_{bc}}=1$. Then since $s_b^2=1$, 
it follows that $s_b(s_bs_c)^{o_{bc}}s_b=s_b^2
=1=(s_cs_b)^{o_{bc}}$, showing that $o_{cb}=o_{bc}$. Moreover,  $s_b^2=1$ implies $o_{bb}=1$. A group generated
by these {\it `simple reflections'} $s_b$, obeying the conditions  $o_{bb}=1$ and $o_{bc}=o_{cb}\ge 2$ for $b\not=c$,
is called a {\it Coxeter group} \cite{coxeter}.  The number of $s_b$'s is the {\it rank} of the group. 

 A Coxeter group of rank $n$ can be 
conveniently represented by
a {\it Coxeter graph} with $n$ nodes, each of which corresponds to a generator $s_b$ of the group. 
 A  line is drawn connecting the pair of nodes $b$ and $c$ provided $o_{bc}\ge 3$,
 with the number $o_{bc}$ written above the line if $o_{bc}>3$.

If there is no line directly connecting node $b$ and node $c$, then 
$(s_bs_c)^2=1$, which implies $s_bs_c=s_cs_b$ because $s_b^2=s_c^2=1$. Thus two
simple reflections  not directly connected mutually commute. If a Coxeter graph is disconnected, then every node
in one part commute with every node in a disconnected part, so the Coxeter group is a direct product of as many
Coxeter subgroups as there are disconnected parts.

All  finite connected Coxeter groups are known, with most of them being  Weyl groups of semisimple Lie algebras. The set
of  roots of
a simple Lie algebra $L$ of rank $n$ is invariant under  reflections about
the hyperplane perpendicular to every simple root. The group generated by these $n$  reflections is known
as the Weyl group of the algebra, and is denoted by $W(L)$. Every Weyl group is a Coxeter group, with the simple
reflections being the generators $s_b$ of the Coxeter group.
If $L$ is expressed as a Dynkin diagram, then the Coxeter graph of $W(L)$
is given by the same Dynkin diagram, with single bonds in the Dynkin diagrams corresponding to $o_{bc}=3$ in the Coxeter
graph, double bonds to $o_{bc}=4$, and triple bonds to $o_{bc}=5$. The arrows do not matter so
$W(B_n)=W(C_n)$. Weyl groups for semisimple Lie algebras are direct product of Weyl groups of simple
Lie algebras.

 In the literature, $W(L)$ is often written simply as $L$, a convention we will adopt here. 
Thus, unless otherwise stated, $A_n$ in this paper is not the Lie group  $SU(n+1)$, nor the finite simple group $A_n$,
nor the alternating group consisting of even permutation of $n$ objects. It is the Weyl group $W(A_n)$. In this notation,  the 
possible Weyl groups are $A_n, B_n=C_n, D_n,  G_2, F_4, E_6, E_7, E_8$, with the subscript
indicating the rank of the Coxeter group.  In particular, it should be noted that $A_1$
is simply the cyclic group $Z_2=S_2$, and $A_n$ is the symmetric group $S_{n+1}$.

Other than the Weyl groups, the Dihedral groups $Dih(n)$ are rank-2 finite Coxeter groups,
denoted by $I_2(n)$. The only other finite Coxeter groups are $H_3$ and $H_4$, of ranks 3 and 4 respectively. 
Their Coxeter graphs are both tree graphs, with
$(o_{12},o_{23})=(6,3)$ for $H_3$, and 
 $(o_{12},o_{23},o_{34})=(6,3,3)$ for $H_4$.

Let us now return to $SO(10)$ and its possible
horizontal groups, generated by $n$ $s_b$'s. 
Since there are three generations of fermions, we only need to consider those groups with
three-dimensional irreducible representations (3dIR). These are $A_3, B_3, H_3, B_4, D_4$, whose Coxeter diagrams
are shown in Fig.~1, and no more.
In terms of the Small Group (SG) designations in the GAP
library \cite{gap, rep}, these groups are $A_3=SG([24,12])=S_4,\  B_3=SG([48,48])=Z_2\x S_4,\  H_3=SG([120,35])=Z_2\x `A_5',\ B_4=SG([384,5602])$, and $D_4=SG([192,1493])$.
In these expressions, $S_4$ is the group of permutation of 4 objects, and $`A_5'$ is the group of even permutation of 5 objects.

The Coxeter graphs for these five groups are given in Fig.~1, with the number of lines between $b$ and $c$ equal to $o_{bc}-2$.

\vskip1cm

\begin{picture}(30,20)
\put(3,3){\circle{15}}
\put(10.5,3){\line(1,0){30}}
\put(0,0){1}
\put(48,3){\circle{15}}
\put(55.5,3){\line(1,0){30}}
\put(45,0){2}
\put(93,3){\circle{15}}
\put(90,0){3}
\put(40,-20){ $\mathbf{A_3}$}

\put(153,3){\circle{15}}
\put(160.5,1.5){\line(1,0){30}}
\put(160.5,4.5){\line(1,0){30}}
\put(150,0){1}
\put(198,3){\circle{15}}
\put(205.5,3){\line(1,0){30}}
\put(195,0){2}
\put(243,3){\circle{15}}
\put(240,0){3}
\put(190,-20){ $\mathbf{B_3}$}

\put(303,3){\circle{15}}
\put(310.5,1.5){\line(1,0){30}}
\put(310.5,4.5){\line(1,0){30}}
\put(310.5,-1.5){\line(1,0){30}}
\put(310.5,7.5){\line(1,0){30}}
\put(300,0){1}
\put(348,3){\circle{15}}
\put(355.5,3){\line(1,0){30}}
\put(345,0){2}
\put(393,3){\circle{15}}
\put(390,0){3}
\put(340,-20){ $\mathbf{H_3}$}

\put(23,-83){\circle{15}}
\put(30.5,-84.5){\line(1,0){30}}
\put(30.5,-81.5){\line(1,0){30}}
\put(20,-86){1}
\put(68,-83){\circle{15}}
\put(75.5,-83){\line(1,0){30}}
\put(65,-86){2}
\put(113,-83){\circle{15}}
\put(110,-86){3}
\put(120,-83){\line(1,0){30}}
\put(155,-86){4}
\put(158,-83){\circle{15}}
\put(85,-110){ $\mathbf{B_4}$}

\put(293,-63){\circle{15}}
\put(290,-66){1}
\put(290,-106){4}
\put(293,-103){\circle{15}}
\put(300.5,-63){\line(3,-2){30}}
\put(300.5,-103){\line(3,2){30}}

\put(338,-83){\circle{15}}
\put(345.5,-83){\line(1,0){30}}
\put(335,-86){2}
\put(383,-83){\circle{15}}
\put(380,-86){3}

\put(330,-110){ $\mathbf{D_4}$}

\end{picture}

\vskip4cm
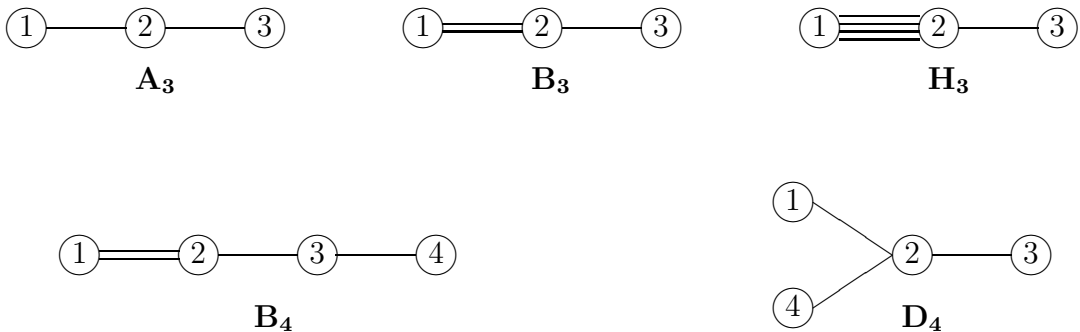
\bc Fig.~1\quad Ranks 3 and 4 finite Coxeter groups with a 3dIR\ec

\section{Three-dimensional irreducible representations}
There are respectively 2, 4, 4, 4, 6 inequivalent 3dIR for $A_3, B_3, H_3, B_4, D_4$ \cite{gap, rep},
but only half of them need to be considered for the following reason. If $\{s_b\}$ is a 3dIR of the fundamental
reflections of a  Coxeter group, then so is $\{-s_b\}$. In three dimensions, these two sets have opposite
determinants, so they cannot be equivalent. However, the constraint imposed on $M$ by $s$ through the relation $s^TMs=M$  
is the same as the constraint imposed by $-s$, hence half the representations do not give rise to anything new. In
what follows we will choose the representation where $\det(s_1)=+1$.

In $A_3, H_3, D_4$, $s_1$ and $s_2$, as well as $s_2$ and $s_3$, are connected by a single bond, hence $(s_1s_2)^3=1$
and $(s_2s_3)^3=1$. Thus $\det(s_1)=+1$ implies $\det(s_2)=+1$ and $\det(s_3)=+1$. This is not necessarily so for $B_2$
and $B_4$, whose $\det(s_3)$ could have either sign.

Another feature of the simply connected diagrams $A_3, H_3, D_4$ is that none of the simple reflections $s_i$ may be the
identity matrix $1$. For example, if $s_1=1$, then $(s_1s_2)^3=s_2^3=1$. Together with $s_2^2=1$, it yields $s_2=1$. 
Similarly $s_3=1$, so this representation is reducible. For that matter, $s_1=1$ or $s_3=1$ is not allowed in $B_3$
either because the rank-2 graph with this node stripped off has no 3DIR, so the  three-dimensional
representation of $B_3$ with $s_1=1$ or $s_3=1$ is not irreducible either. In fact, the only node where 1 is allowed is $s_1$
in $B_4$, and the only 3DIR are those with $s_1=\pm 1$ and $s_2, s_3, s_4$ form a 3DIR of $A_3$.

Since $s_1$ and $s_3$
are not directly connected in the Coxeter graphs, they commute so they can be diagonalized simultaneously.
For the rank-3 groups, neither of them can be 1, nor is $s_1=s_3$ allowed, for otherwise the representation
is essentially the same as a rank-2 group with $s_3$ stripped, whose three-dimensional representation
is reducible. For $A_3$ and $H_3$, it is thus possible to choose a basis so that $s_1=\diag(-1,-1,+1):=x$, and $s_3=\diag(+1,-1,-1):=z$.
For $B_3$, we can choose $s_1=x$ but $s_3$ may be $z$ or $-z$.
The remaining simple reflection $s_2$ is determined by the conditions $(s_is_2)^{m_{i2}}=1$ and the result is shown in Table 1 and Eq.~(1).
 The number $\varphi:=(1+\sqrt{5})/2$ 
is the golden ratio, with $\varphi^{-1}=\varphi-1=(\sqrt{5}-1)/2$.

$$\ba{|c|c||c|c|c|c|}\hline
{\rm group}&{\rm IR}&s_1&s_2&s_3&s_4\\ \hline\hline
A_3&1&x&y_1&z&-\\ \hline
B_3&1&x&y_2&z&-\\
&2&x&y_3&-z&-\\ \hline
H_3&1&x&y_4&z&-\\
&2&x&y_5&z&-\\  \hline
B_4&1&1&x&y_1&z\\
&2&1&-x&-y_1&-z\\ \hline
D_4&1&x&y_1&x&z\\ 
&2&x&y_1&z&x\\
&3&x&y_{1}&z&z\\
\hline\ea$$
\bc Table 1. Irreducible representations (IR) of the five finite Coxeter groups\ec

Their detailed matrix forms are:
\be
y_1&=&{1\over 2}\bm{ -1&\rd&-1\cr \rd&0&-\rd\cr -1&-\rd&-1}\em,\quad
y_2={1\over 2}\bm{ -1&1&\rd\cr 1&-1&\rd\cr \rd&\rd&0}\em,\quad
y_3={1\over 2}\bm{ 1&-1&\rd\cr -1&1&\rd\cr \rd&\rd&0}\em\nn\\  \nn\\ 
y_4&=&{1\over 2}\bm{-1&-\varphi^{-1}&\varphi\cr -\varphi^{-1}&-\varphi&-1\cr \varphi&-1&\varphi^{-1}}\em,\quad
y_5={1\over 2}\bm{-1&\varphi&-\varphi^{-1}\cr \varphi &\varphi^{-1}&-1\cr -\varphi^{-1}&-1& -\varphi}\em
\ee

For the rank-4 groups, as remarked earlier, $B_4$ is obtained from the $A_3$ representation with a $s_1=\pm 1$ attached. For $D_4$, it 
collapses into an
$A_3$ with either $s_1, s_3$, or $s_4$ removed. With $s_2$ given by that in $A_3$, $s_1$ fixed to be $a$, then $(s_3,s_4)$
must be either $(x,z), (z,x)$, or $(z,z)$. These remarks about $B_4$ and $D_4$ have been incorporated in Table 1.

\section{Constraint on Fundamental Mass Matrices}
The general forms of a symmetric and an antisymmetric mass matrix are
\be M_s:=\bm{a&b&c\cr b&d&e\cr c&e&f\cr}\em,\quad M_a:=\bm{0&\beta&\gamma\cr -\beta&0&\epsilon\cr -\gamma&-\epsilon&0}\em\labels{mat}\ee
Table 2 gives the relations imposed on their parameters by the symmetry relation $s^TMs=M$ for each of the $s$ in Table 1.
If $s=1$, then there is no constraint whatsoever, enabling $M_s$ to be any complex symmetric matrix and $M_a$ to be any complex antisymmetric matrix.
$$\ba{|c|c|c|}\hline
s&M_s&M_a\cr\hline\hline
1&-&-\cr
x&c=e=0&\gamma=\epsilon=0\cr
z&b=c=0&\beta=\gamma=0\cr
y_1&c=-d+(a+f)/2,\ e=-b-(a-f)/\rd&\gamma=\rd\beta,\ \epsilon=-\beta\cr
y_2&b=f-(a+d)/2,\ c=e-(a- d)/\rd&\gamma=-\beta/\rd,\ \epsilon=\beta/\rd\cr
y_3&b=f-(a+d)/\rd,\ c=e+(a-d)/\rd&\gamma=\beta/\rd,\ \epsilon=-\beta/\rd\cr
y_4&c=b+[-(\varphi+\varphi^{-1})a+(\varphi^{-1}-1)d+(\varphi+1)f]/2&\gamma=(1-\varphi^{-1})\beta,\ \epsilon=\varphi^{-1}\beta\cr
 & e=-b+[\varphi^{-1}a+d-\varphi f]/2  &\cr
y_5&c=b+[-(\varphi+\varphi^{-1})a-(\varphi+1)d-(\varphi^{-1}-1)f]/2&\gamma=(1+\varphi)\beta,\ \epsilon=-\varphi\beta\cr
&e=-b+[-\varphi a+d+\varphi^{-1} f]/2  &\cr
\hline\ea$$
\bc Table 2. Symmetry constraints on symmetric $M_s$ and antisymmetric $M_a$ mass matrices\ec

\section{Fermion Mass Matrices}
Since every fermion is contained in {\bf 16}, the fermion mass matrices $m_\a\ (\a=u, d, e, \nu)$ can  be obtained from the 
$16\x 16$ fundamental mass matrices
$M^{10}\equiv H,\  M^{120}\equiv G$ and $M^{\overline{126}}\equiv F$. 
$H$ contributes equally to $m_\nu$ and $m_u$, and equally to $m_e$ and $m_d$, whereas
$F$ contributes $-3$ times as much to $m_\nu$ as $m_u$, and $-3$ times as much to $m_e$ as $m_d$. Only {\bf $\overline{126}$} contains a Standard-Model singlet, so the Majorana mass matrices receives a contribution only from
$F$. The effective mass matrix for the active neutrinos is obtained from
the neutrino Dirac mass matrix $m_\nu$ and the Majorana mass matrices $m_R$ and $m_L$ by the formula
\be \overline m_\nu=-m_\nu m_R^{-1}m_\nu^T+m_L,\ee
where the first term comes from the type-I seesaw mechanism and the second term comes from the type-II seesaw. 

These relations between fermion mass matrices and fundamental mass matrices are summarized in Table 3, where $r_i$ are 
arbitrary coefficients. The normalization of $H, F$ and $G$ is determined by choosing the coefficients of $m_d$ in all of them to be 1.
$$\ba{|c|ccc|}\hline
&H&F&G\\ \hline
m_u&r_1&r_2&r_3\\
m_d&1&1&1\\
m_e&1&-3&r_4\\
m_\nu&r_1&-3r_2&r_5\\ \hline
m_R&0&r_6&0\\
m_L&0&r_7&0\\ \hline \ea$$
\bc Table 3. Relations between fermion and fundamental mass matrices\ec

There are currently 18=13+5 experimentally measured values associated with the fermion mass matrices, in which 5 are neutrino
quantities and 13 are non-neutrino. The neutrino ones are the two oscillation mass gaps, and the three PMNS mixing angles.
The others are the nine charged-fermion masses and the four CKM mixing parameters.

In general, both the fundamental mass matrices $H, F, G$ and  the coefficients $r_i$ are complex,
though phases may be chosen to render one $r_i$ per fermion mass matrix real.  Together they contain many more parameters
 than the available experimental quantities, so various ways have been devised in the literature \cite{so10, so10hs, moh,ft} to reduce the number of parameters
to be close to the experimental number of 18. Dropping all contributions from $G$ is one way.
Another way is to assume the fundamental and the fermion mass matrices to be hermitian,  hence all the coefficients $r_i$ to be
real. This assumption can be justified if CP symmetry is broken spontaneously, a theory  sometimes referred to as 
the charge-conjugation-conservation (CCC) \cite{CCC} theory. Since $H$
and $F$ are hermitian and symmetric, their matrix elements are real, thus each is described by 6 (real) parameters. $G$ is hermitian and 
antisymmetric, hence its matrix elements are purely imaginary, with 3 parameters.  From Table 3, we see that there are 7 $r_i$'s (6 if
only one of type-I and type-II seesaw is present). Altogether there are 22 parameters, still larger than the 18 available experimentally,
thus more constraints can be imposed.

Horizontal symmetry is another way to reduce the number of parameters \cite{so10hs}. With a built-in  finite symmetry, it must 
 be either $A_3, B_3, H_3, B_4$, or $D_4$. Each fundamental mass matrix $M$ must be invariant under a simple reflection
generator $s$ of the group, but its $SO(10)$ transformation property is up to us to choose. For example,
for rank-3 groups, we can assign the three of them to 
transform like $H, F, G$ respectively, or we can assign two of them to transform like $H$, and one like $F$, etc. 
Since there are two constraints per simple reflection, in the first case we reduce the total parameters of $H, F, G$ from
15 to 9, yielding a total of 16 parameters in a CCC theory, two short of the experimental quantities. 
If that fits well, it is a strong indication of the validity of the horizontal symmetry.
For rank-4 groups, at least two of $M$'s must have the same $SO(10)$ transformation property,
which tends to increase the number of available parameters compared to the rank-3 groups,  but what that 
is depends on the details.
 
All in all,  there are many ways to assign the horizontal and vertical transformation properties of
the fundamental mass matrices, thereby producing many possible models even for a single horizontal group. 
A systematic attempt to cover all possibilities  involves a large amount of work, but the amount is finite
because there are only five possible groups. For each fit, we must use experimental values extrapolated to GUT energy,
and that depends on the detailed dynamics in between, which further adds to the complication. 
Since the five horizontal symmetries are built into $SO(10)$ and theoretically derived, it is hoped that the constraints they
provide would be better than those without a strong theoretical basis. However,
we will not attempt any of these fits
in the present article.

It should be mentioned that in the discussion above, 
we implicitly assumed that every $M$ has a single $Z_2$ symmetry. Recall however that the symmetry
could be $Z_2\x Z_2$. In that case there would be three or four constraints for the matrix elements of $M$, rather
than just two. 

In the opposite direction, we may assign  two $M$'s with the same $SO(10)$ transformation
to be invariant under different simple reflections, $s_i$ and $s_j$. The end result is like
having only one $M$, but  with fewer constraints on its matrix elements. For example, if $M_x$ is invariant under $x$
of Table 1 and $M_z$ is invariant under $z$, and both are of  type $H$, then their sum is still of  type $H$,
and according to Table 2 it is of the form
\be
M:=M_x+M_z=\bm{a&b&0\cr b&d&0\cr 0&0&f\cr}\em+\bm{a'&0&0\cr 0&d'&e'\cr 0&e'&f'}\em.\ee
The result is a symmetric matrix with the (13) and (31) elements zero, and no further
constraint on any of the other matrix elements.  As another example, consider $M=M_{y_1}+M_{z}$. Then
\be
M:=M_{y_1}+M_{z}=\bm{a&b&c\cr b&d&e\cr c&e&f\cr}\em +\bm{a'&0&0\cr 0&d'&e'\cr 0&e'&f'}\em,\ee
where $c=-d+(a+f)/2$ and $e=-b-(a-f)/\rd$. The result is a symmetric matrix with no constraint whatsoever on any of its elements.
The same would be true for the sum $M=M_{y_i}+M_z$ for $i=2,3,4,5$.

\section{Horizontal Symmetry and Other Constraints}
To compare the use of built-in horizontal symmetry to impose constraints with other approaches in the literature, we discuss 
 two specific examples in this 
section as an illustration.
\subsection{\bf $S_4$}

An interesting $SO(10)$ model possessing an $S_4$ horizontal symmetry is given in Ref.~\cite{moh}. The fundamental mass matrices in that model are \cite{note}
\be
H=\bm{0&0&0\cr 0&0&0\cr 0&0&\tilde M\cr}\em,\quad H'=\bm{0&\delta&-\delta\cr \delta&0&0\cr -\delta&0&0\cr}\em,\quad
F=\bm{0&m_1&m_1\cr m_1&m_0&m_1-m_0\cr m_1&m_1-m_0&m_0\cr},\em\labels{s4}\ee
and $G=0$,
where $H'$ has the same $SO(10)$ transformation property as $H$.
The parameter $\tilde M$ is real, and $\delta, m_0, m_1$ are complex. 

Since $A_3=S_4$ is one of the five  built-in symmetries, the success of this model  seems to confirm their presence.
 Unfortunately this is not so because the residual symmetry left behind after the breaking of 
the $S_4$ in Ref.~\cite{moh} is not
the simple reflection generators $s_1, s_2, s_3$ of $A_3$. Thus whether the built-in $A_3$ is a
symmetry or not must be decided by a new fit. 

To see this point in more detail, let us express the generators $s_1=x,\ s_2=y_1,\ s_3=z$ of Table 1 in a basis that gives rise to
$F$ in \eq{s4}. This is accomplished by making a similarity transformation using
\be U={1\over\rs}\bm{\rd&2&0\cr \rd&-2&\rt\cr \rd&-1&-\rt\cr}\em,\ee
to get the generators $x'=UxU^T,\  y_1'=Uy_1U^T$, and  $z'=UzU^T$ in the new basis:
\be x'=-\bm{1&0&0\cr 0&0&1\cr 0&1&0\cr}\em,\ y_1'={1\over 4}\bm{2&-\rs&\rs\cr -\rs&-3&-1\cr \rs&-1&-3}\em,\
z'={1\over 3}\bm{-1&2&2\cr 2&-1&2\cr 2&2&-1\cr}\em.\ee
The invariant conditions $s^TM_ss=M_s$ for the symmetric matrix $M_s$ in \eq{mat} can be worked out to be:
\bn
\i for $s=x'$:\quad $c=b,\ f=d$; 
\i  for $s=y_1'$:\quad $c=\rs(d-a-e)-5b,\ f=6(a+e)-5d+4\rs b$;
\i  for $s=z'$:\quad $c$ and $f$ are determined
by the conditions $a+b+c=b+d+e=c+e+f$.
\en
 In other words, $M_{x'}$ is 2-3 symmetric and $M_{z'}$ is magic. Since $F$
in \eq{s4} is 2-3 symmetric and magic, it can be obtained either from $M_{x'}$ or $M_{z'}$. The other two $M$'s must then
be equal to $H$ and $H'$ in \eq{s4}, if the $S_4$ breaking in \cite{moh} respects the residual invariants of $x', y_1', z'$.
There is no problem in getting $H$ but it is not possible to get $H'$. This shows that the residual symmetry for \eq{s4}
is not $x', y_1', z'$.
\subsection{Fritzsch Texture}
We saw in the last section that the CCC theory contains 22 parameters, still more than the 18 experimental quantities available. 
One way to reduce the parameters further is to assume every fundamental and fermion mass matrices to have the Fritzsch texture
\cite{fr}. That is, to assume not only that they are hermitian, but also that their (11), (13), and (31) matrix elements vanish. 
This cuts down 5 more parameters to a total of 17, one short of the experimental quantities. Reasonable fits are reported
in such a scheme \cite{ft}.

We saw in the last section that the CCC theory with built-in horizontal symmetry groups of rank-3 has 16 parameters. If we go to
rank-4 groups, then the number of parameters increase. For example, in $D_4$, if we assign the fourth mass matrix
to be a $G$-type, then two more parameters are added to make it 18: one in the matrix element of the new $G$, and one each
from the coefficients for $m_u$ and $m_\nu$ for this new $G$. If we assign the fourth fundamental mass matrix to be of
$H$ or $F$ type, then even more free parameters are available. If we use $B_4$, since  $s_1=1$ does not place any
constraint on the fundamental mass matrices, there are more parameters still. All in all, there seems to be a sufficient
number of parameters to make a successful experimental fit quite possible.

There is some formal similarity between the horizontal-symmetry constraints in Tables 1 and 2 on the one hand,  
and the Fritzsch-texture constraint on the other.
First of all, both place two constraints on each of the fundamental mass matrices. Secondly, we see in Table 1 that $x$ and $z$
are common generators for all the groups. If we let $M_x$ to be $H$ type and $M_z$ to be $F$ type, then we see in Table 2
that both their (13)=(31) matrix elements vanish, just like in the Fritzsch texture. In addition, for $M_x$, instead of having (11) zero
as in the Fritzsch texture,
it is (23)=(32) that is zero. For $M_z$, instead of having (11) zero, it is (12)=(21) that is zero.  Moreover, in the case of $D_4$,
we can always assign another $x$ or $z$ to $G$ to make its (13) element $\gamma$ vanish as well, as in Fritzsch texture.
The main difference with the Fritzsch texture is that the zeros of the latter are in fixed positions for all fundamental 
matrices, but that is not
the case for the built-in symmetries. 

\section{Summary}
The main purpose of this article is to point out that there is a built-in horizontal symmetry for $SO(10)$, in the form
of a Coxeter group. For general fundamental mass matrices without any constraint, that Coxeter group is infinite in size. 
If we demand the  symmetry group
to be finite, then it is limited to only five groups. 
This result is based on the reasonable assumption that natural symmetries are the residual symmetries
left behind after breaking, an assumption
already used fairly widely in analyzing neutrino physics.

Some immediate consequences of this conclusion are discussed. This includes how the constraints from such horizontal symmetries
reduce the number of free parameters used to fit the data. The details of these constraints are quite different from those used
in the literature. This point is illustrated in the last section in an $S_4$ model, and for the Fritzsch texture.

Since  finite built-in horizontal symmetries for $SO(10)$ can be derived, it is hoped that they can offer better constraints
than those without a theoretical basis. However, at present that remains only a hope because 
no attempt has been made  to fit the data in this article. This important task of fitting will be left to future research.

I am indebted to J. Bjorken, H. Fritzsch,  H.J. He, W. Liao, Y. Mimura, W. Rodejohann, and J. Stembridge for helpful discussions.


\begin{thebibliography}{9}
\bibitem{note} See eqs.~(7)-(9) of the last paper in Ref.~\cite{moh}. The parameter $M$ is replaced by $\tilde M$
here so as not to be confused with the symbol for the fundamental mass matrix.
\end{thebibliography}
\end{document}